\documentclass[12pt,preprint]{aastex}
\shorttitle{Roche Problem}
\shortauthors{Seidov }

\begin{document}
\title{The Roche problem: some analytics }
\author{Zakir F. Seidov}
\affil{Research Institute, College of Judea and Samaria, Ariel 44837,Israel}
\email{zakirs@yosh.ac.il}
\date{\today}
\begin{abstract}
Exact analytical formulae are derived for
the potential and mass ratio as a function of Lagrangian points
position, in the classical Roche model of the close binary stars.
\end{abstract}
\section{Introduction}
The Roche model is widely used in interpretation of the close binary
star observations. Several authors derived the various approximations
to solve the Roche problem and presented numerical tables, see e.g.
\citet{Plav64,KiTo70,Eggl83,Moch84,Morr85,Morr94}.
 Not intending to solve the whole problem analytically,
I rather show here the way how some analytical relations can be found.
The  idea is to reverse the problem: instead of finding, e.g., the first
Lagrangian point, $x_1$,
as function of binary mass ratio $q$ we seek solution for $q$ as function of $x_1$.
\section{Basic equation}
The  basic equation of the classic Roche problem is
the formula for surfaces of the (primary) star as equipotential :
\begin{equation}
\Psi(x,y,z)= \left( x - \frac{q}{1 + q}
           \right)^2 + y^2 + \frac{2\,q}
   {\left( 1 + q \right) \, {\sqrt{{\left( -1 + x \right) }^
          2 + y^2 + z^2}}} + \frac{2} {\left( 1 + q \right) \,
     {\sqrt{x^2 + y^2 + z^2}}}.%=\Psi_0.
\label{basic} \end{equation}
I use in eq. (\ref{basic}) notations of \citet{Moch84}: the $x$ axis is aligned
along the stars' centers, $z$ axis is parallel to the rotation axis,
$\Psi$ is "normalized" potential, in units of $G(M_1 + M_2)/(2\,A)$,
$q=M_2/M_1<1$, and $x, y, z$ are in units of $A$, the distance between the centers of
binary components with masses $M_1$ and $M_2$.
As I here consider only the first (inner) and second (outer)
Lagrangian points problem, it is sufficient to consider the eq. (\ref{basic})
at the $x$ axis.
\section{The first Lagrangian point}\label{LP1}
We write down eq. (\ref{basic}) with $y=z=0,\;0<x<1$, and look for the minimum of the
function:
\begin{equation}
\Psi_1(x)=\frac{2\,q}{\left( 1 + q \right) \,\left( 1 - x \right) } +
\frac{2}{\left( 1 + q \right) \,x} +  {\left(x - \frac{q}{1 + q}\right) }^2.
 \label{psixx11} \end{equation}
at some $x=x_1$, with $0<x_1<1$.
The important observation, from eq. (\ref{psixx11}),
is that $\Psi_1(q,x)=\Psi_1(1/q,1-x)$, if $0<x<1$
(not in general case!).
From eq. (\ref{psixx11}), we find the condition $d\,\Psi_1(x)/d x=0$,
which we rewrite after some algebra as function of $q(x_1)$:
\begin{equation}
q(x_1)=\frac{{\left( 1 - x_1 \right) }^3\,\left( 1 + x_1 + x_1^2 \right) }{x_1^3\,
    \left( 3 - 3\,x_1 + x_1^2 \right) }.
  \label{qvsx1} \end{equation}
We notice, from eq. (\ref{qvsx1}), the  elegant relation (also having a clear physical
 meaning) $q(x_1)\,q(1-x_1)=1$.
We underline that  eq.~(\ref{qvsx1}) gives the fully analytical and {\em exact}
 relation between values of $q$ and
of the first Lagrangian point $x_1$. See, e.g. table 1 in
\citep{Moch84}, where $Q$ and $X1$ stand for our $q$ and $x_1$,
respectively,
%It's worth noting that all digits in the table 1 in \citep{Moch84} are correct
and compare the difference in difficulty of calculations
by method of  \citet{Moch84} and by formulas
(\ref{qvsx1},\ref{psivsx1},\ref{qvsx2},\ref{psivsx2}) of this note.

Now in order to find value  of the potential corresponding to the first  Lagrangian
point, we may use the eqs.  (\ref{psixx11}) and  (\ref{qvsx1})  together, or, in the
spirit of this note, exclude $q$ from eqs.  (\ref{psixx11}) and (\ref{qvsx1})
and find explicit function $\Psi_1(x_1)$:
\begin{equation}
\Psi_1(x_1)=
%\frac{3 - 12\,x_1 + 27\,x_1^2 - 40\,x_1^3 +  41\,x_1^4 - 14\,x_1^5 -14\,x_1^6 +
%16\,x_1^7 - 4\,x_1^8}{{\left( 1 -  2\,x_1 + x_1^2 + 2\,x_1^3 - x_1^4 \right) }^2}.
\frac{3 - 12\,t + 15\,t^2 - 10\,t^3 - 4\,t^4}{{\left( -1 + 2\,t + t^2 \right)
}^2};\;t=x_1(1-x_1).
\label{psivsx1} \end{equation}
 We introduced the additional variable $t$ into Eq.~(\ref{psivsx1}) in order
  to explicitly show that $\Psi_1(x_1)=\Psi_1(1-x_1)$ if  $0<x_1<1$
(not in general case!).
 Eq.~(\ref{psivsx1}) gives the fully analytical and {\em exact} relation between the
 values of the first Lagrangian point $x_1$
  and of the corresponding potential $\Psi_1(x_1)$.
  See, e.g. table 1 in
\citep{Moch84}, where $C1$ and $X1$   stand for our $\Psi_1(x_1)$ and
$x_1$, respectively.
\section{The second Lagrangian point}\label{LP2}
Now we look for minimum of the function
(note the difference from eq. [\ref{psixx11}]):
\begin{equation}
\Psi_2(x)=\frac{2\,q}{\left( 1 + q \right) \,\left( x - 1 \right) } +
\frac{2}{\left( 1 + q \right) \,x} + {\left(x - \frac{q}{1 + q}\right) }^2,
 \label{psixx22}\end{equation}
at some $x=x_2$, with $1<x_2<2$.
Repeating the procedure of the section \ref{LP1}, we get the solution
for $q$ as function of $x_2$:
\begin{equation}
q(x_2)=\frac{{\left(x_2 - 1  \right) }^3\,
 \left( 1 + x_2 + x_2^2 \right) }{x_2^2\,(2-x_2)\,
 \left( 1 - x_2 + x_2^2  \right) }.
 \label{qvsx2} \end{equation}
We notice that by contrast with eq. (\ref{qvsx1}),
it is not evident from eq. (\ref{qvsx2}) that we have
the relation  $q(x_2)\,q(1-x_2)=1$, because here $x_2>1$, but $(1-x_2)<0$, while the
eq.~(\ref{qvsx2}) is derived under condition $1<x_2<2$; in the case of
 eq.~(\ref{qvsx1}), both $x_1$ and  $(1-x_1)$ were in the (open) interval (0,1)).

 To prove the validity of the relation $q(x_2)\,q(1-x_2)=1$, we should return
 to the basic equation (\ref{basic}), put there $y=z=0, x<0$,
 and look for the minimum of the function:
\begin{equation}
\Psi_3(x)=\frac{2\,q}{\left( 1 + q \right) \,\left( 1 - x \right) } -
 \frac{2}{\left( 1 + q \right) \,x} + {\left(x - \frac{q}{1 + q}\right) }^2,
 \label{psixx33} \end{equation}
at some $x=x_3$, with $x_3<0$.
Note the differences between functions $\Psi_1(x), \Psi_2(x)$, and $\Psi_3(x)$.
Repeating above procedure for $\Psi_3(x)$, we get:
\begin{equation}
q(x_3)=\frac{\left( 2 - {x_3}\right)\,%
x_3^2\,\left( 1 - {x_3}+x_3^2 \right) }{{\left(%
{x_3}-1 \right) }^3\,\left( 1 + {x_3} + x_3^2\right)}.
\label{qvsx3} \end{equation}
Now, from eqs. (\ref{qvsx2}) and (\ref{qvsx3}), we have $q(x_2)\,q(x_3)=1$,
if $x_3=1-x_2$, {\em and if} $x_2>1,\, x_3=1-x_2<0$, QED.
Eq.~(\ref{qvsx2}) (together with eq.~[\ref{qvsx3}])
 gives the fully analytical and {\em exact} relation between $q$ and
the value of the second Lagrangian point $x_2$. See, e.g. table 1 in
\citep{Moch84}, where $Q$ and $X2$   stand for our $q$ and $x_2$, respectively.

Now in order to find value  of the potential corresponding to the second  Lagrangian
point, we may use the eqs. (\ref{psixx22}) and  (\ref{qvsx2}) together, or,
 again in the spirit of this note, exclude $q$ from eqs. (\ref{psixx22})
 and  (\ref{qvsx2}) and find explicit function $\Psi_2(x_2)$:
\begin{equation}
\Psi_2(x_2)=\frac{-1 - 4\,x_2 + 27\,x_2^2 - 36\,x_2^3 +
 9\,x_2^4 + 18\,x_2^5 - 14\,x_2^6 + 4\,x_2^7}{{\left( -1 + 2\,x_2 + x_2^2 -
2\,x_2^3 + x_2^4 \right) }^2}.
  \label{psivsx2} \end{equation}
 Eq.~(\ref{psivsx2}) gives the fully analytical and {\em exact} relation between the
 values of the second Lagrangian point $x_2$
 and of the corresponding potential $\Psi_2(x_2)$. See, e.g. table 1 in
 \citep{Moch84}, where $C2$ and $X2$   stand for our $\Psi_2(x_2)$ and
$x_2$, respectively.
\section{Summary}
In this short note we present the exact analytical relations for the first
and second Lagrangian point of the classical Roche problem.
In practice, the "exact" formulas are not necessarily the most convenient  ones,
hence  different approximate expressions in literature (see \citep{
Kopa59,Plav64,KiTo70,Eggl83,Moch84,Morr85,Morr94}, among others), and some relevant
approximate formulas will be given elsewhere.
Still, the exact formulas have their own beauty and
are more relevant as solutions to the classical problems such as the problem
by  \citet{Roch47}.
%\acknowledgments

My thanks are due to anonymous referee for encouraging criticism.

\end{document}